\documentclass[12pt]{iopart}
\pdfoutput=1

\usepackage{color}
\usepackage {graphicx}
\def\>{\right\rangle}
\def\<{\left\langle}
\def\be{\begin{equation}}
\def\ee{\end{equation}}
\def\ba{\begin{array}{lll}}
\def\ea{\end{array}}

\def\beq{\begin{eqnarray}}
\def\eeq{\end{eqnarray}}

\newcommand{\D}{\mathcal{D}}

\def \ket#1{\mathinner{|{#1}\rangle}}
\def \bra#1{\mathinner{\langle{#1}|}}

\begin{document}
\title{Computation of transient dynamics of energy power for a dissipative two state system}

\author{{\bf M Carrega$^{1}$, P Solinas$^{1}$, A Braggio$^{1}$, and M Sassetti$^{2,1} $}}
\address{$^1$ SPIN-CNR, Via Dodecaneso 33, 16146 Genova, Italy}
\address{$^2$ Dipartimento di Fisica, Universit\`a di Genova, Via Dodecaneso 33, 16146 Genova, Italy}
\ead{}
\vspace{10pt}
\begin{abstract}
We consider a two level system coupled to a thermal bath and we investigate the variation of energy transferred to the reservoir as a function of time. The physical quantity under investigation is the time-dependent quantum  average power. 
 We compare quantum master equation approaches with functional influence method.
 Differences and similarities between the methods are analyzed, showing deviations at low temperature between the functional integral
approach and the predictions based on master equations.
\end{abstract}
\section{Introduction}
Recently a great interest has been devoted to the study of thermal properties of very small systems \cite{esposito2009nonequilibrium, campisi2011colloquium, pekola2015}. 
When the system dimensions shrink, the classical concepts of thermodynamics have to be reconsidered carefully. Indeed, dealing with system at the submicron or nanometer scale, fluctuations and quantum effects become important and, eventually, play a crucial role \cite{esposito2009nonequilibrium, campisi2011colloquium, talkner2007fluctuation}. 
The field of ``quantum thermodynamics'' aims exactly to the investigation of energy exchanges from both a fundamental and an applicative point of view.
On one hand, the extension of classic concepts to their quantum counterparts is still an open problem. Indeed, heat and work cannot be interpreted as quantum observables and precise protocols have to be considered to deal with these quantities \cite{campisi2011colloquium, talkner2007fluctuation, campisi2009fluctuation, solinas2013w}.
On the other hand, a deep understanding of these concepts can serve as a starting point to build new devices, such as heat engines or thermal diodes, potentially increasing the efficiency of quantum machines \cite{segal2005, segal2013, saito2007, vinkler2014, ojanen2014, saito2013, ludovico2014}. In fact, concepts as work and heat are well defined in macroscopic classical system but at quantum level different approaches must be introduced \cite{campisi2011colloquium, pekola2015, solinas2013w, solinas2015, schmidt2014, gasparinetti2014heat, carreganjp} to properly define these quantities. 

Basically two are the main issues in this framework.
The first is related to the measurement problem and how to extract information about the change of the energy of a quantum system \cite{campisi2011colloquium, talkner2007fluctuation, solinas2013w, dorner2013extracting, mazzola2013measuring, batalho2014, wolfarth2014distribution}.
The second one is the description of dissipation in an open quantum system at very low temperature, where coherent effects dramatically affect the dynamics \cite{gasparinetti2014heat, carreganjp, weiss1999quantum}. 
From the experimental side, recent advancements have been achieved  regarding the study of work in a closed quantum system \cite{dorner2013extracting, mazzola2013measuring, batalho2014}
{ and important progresses in the measurement of the  dissipated energy through an open quantum system to a reservoir have been reported} \cite{pekola2015, pekola2013calorimetric, gasparinetti2014fast}.  

Of particular interest is the study of energy exchange between a quantum system and a single thermal bath (reservoir). Here new questions arise, such as the knowledge of the statistics of dissipated energy in different conditions, when the quantum system is subject to a time-dependent driving field, in presence of strong system-bath coupling and also in the low temperature regime \cite{schmidt2014, gasparinetti2014heat, carreganjp}.\\
{In this article, we study the energy dissipated by a quantum system into the thermal bath.
There are different proposals to measure it including a direct double measure \cite{campisi2011colloquium, campisi2009fluctuation}, a measure of the temperature of the bath along the evolution \cite{pekola2015, pekola2013calorimetric, gasparinetti2014fast} or a measure with a quantum detector through the system degrees of freedom \cite{solinas2013w, solinas2015}.
}
To properly define and to calculate the variation of energy of the reservoir we refer to the so-called two-measurement protocol, which is based in two subsequent strong measurements of the energy of the bath in a given time interval. In this scheme one can characterize all statistical properties of variation of energy from the reservoir perspective \cite{gasparinetti2014heat, carreganjp}. 

Once a protocol has been defined 
, the main problem is to solve the quantum dynamics. Different methods exist to face this problem \cite{weiss1999quantum, ingold2002path, breuer, mandel}.
Different methods are based, for example, on the so-called master equation approach \cite{gasparinetti2014heat, breuer, mandel} or stochastic quantum jump \cite{pekola2015, schmidt2014, frankprl, pekola2013calorimetric}, but they often rely on strong approximations necessary to get simple and computable expressions.
An alternative approach is given in Ref. \cite{carreganjp}, where it  was developed a formalism based on the path-integral approach  \cite{ingold2002path, feynman1963theory, caldeira1983path, leggett1987dynamics, chen1989}, and it was obtained a general expression for the functional which formally embodies all information about the dissipated energy between a reservoir and a generic quantum system.

In this article, we consider the paradigmatic case of a two state system (TSS) coupled to a thermal bath. We focus on the quantum average power in the so-called two-measurement protocol (TMP) and we study its dynamics at different temperatures using different approaches. We postpone to further publications the investigation of the full statistics of energy exchange
{ and focus on the first moments, i.e., the quantum average dissipated heat and power with the functional integral approach \cite{carreganjp}}.
{ These observables are compared} to the similar ones obtained with a generalized master equation approaches \cite{schmidt2014, gasparinetti2014heat, breuer, mandel}. We will show where the results obtained with the latter approaches reproduce at high temperature the same behaviour of the functional integral approach, and they fail in the low temperature limit.

The paper is organized as follows.
In section \ref{sec:model} we  describe the model for the dissipative two state system and we introduce useful notations. In section \ref{sec:approach} we recall the two-measurement protocol , giving the expression of the characteristic function of the energy exchange between the reservoir and the TSS.
We present the  different approaches, giving analytical results for the quantum average power. Finally, section \ref{sec:result} is devoted to the discussion of the main results.
\section{Model and general settings}
\label{sec:model}
We consider a two state system (TSS) described by the Hamiltonian ($\hbar=1$) \cite{weiss1999quantum, breuer, grifoni1998driven, hartmann2000driven} 
\be
\label{eq:hs}
H_S = - \frac{\Delta}{2}\sigma_x - \frac{\epsilon}{2}\sigma_z~,
\ee
where $\Delta$ represents the tunneling amplitude between the two states and $\epsilon$ is an external bias. For sake of simplicity in the following we will restrict the discussion to the case of constant external bias $\epsilon (t)=\epsilon$.
The eigenbasis of $\sigma_z$ are $|l \rangle $ and $|r \rangle$ with eigenvalues $\pm 1$.
The TSS is coupled to an external bath described in terms of an infinite set of harmonic oscillators \cite{weiss1999quantum, breuer, ingold2002path}.
 The total Hamiltonian reads
\be
H= H_S + H_B + H_I~,
\ee
where 
\be
H_{B} = \sum_{\alpha=1} ^N \Big [ \frac{p_\alpha^2}{2 m_\alpha } + \frac{1}{2} m_\alpha \omega^2_\alpha x_\alpha^2 \Big ]  ~,
\ee
 and $H_I$ are the bath and system-bath interaction Hamiltonian, respectively.
{We assume a {\it weak} and linear coupling with
\be
H_I=-\frac{q_0}{2}\sigma_z \otimes \sum_{\alpha=1}^N c_\alpha x_\alpha~,
\ee
where $q_0 $ is a characteristic position of the TSS, $x_\alpha$ a generic position of the $\alpha$-th harmonic oscillator and $c_\alpha$ describes the strength of interactions.
\section{Calculation of energy exchange}
\label{sec:approach}
{Here we focus on the variation of energy of the reservoir in a given time interval $t-t_0$. 
We refer to the so-called two-measurement protocol  of the bath energy \cite{esposito2009nonequilibrium, campisi2011colloquium, gasparinetti2014heat, frankprl, pekola2013calorimetric}.
This scheme consists in an initial projective measurement of the bath Hamiltonian at time $t_0=0$, followed by the free evolution and a second projective measurement at the final time $t$.
By repeating this protocol, one can define the statistical average of the variation of energy of the bath and its whole statistics\cite{gasparinetti2014heat}.\\
The main advantage of measuring the bath energy lies in the absence of perturbation induced by the initial projective measurement on the initial state of the system, without statistically affecting its evolution.
In general, a projective measurement on a quantum system would destroy the off-diagonal elements of the density matrix, which describe coherences between eigenstates, of the measured observable.
This affects the  dynamics of the system, thus modifying also the energy exchange\cite{campisi2011colloquium, gasparinetti2014heat}.
On the contrary, initially the bath can be safely considered to be in a thermal state, described in terms of the equilibrium density matrix $\rho_B(0)=e^{-\beta H_B}/Z_B$, with $Z_B$ the partition function where $\beta =1/(k_B T)$ the inverse temperature.
Therefore, is diagonal in the $H_B$ eigenstates and the initial projective measure has no effect on the dynamics of the averaged quantities \cite{campisi2011colloquium, solinas2015}.
}

Having clarified the motivations behind the reservoir measures, we can now discuss how to obtain the energy exchange statistics.
We assume that initially the total density matrix $\rho$ can be expressed in a product form $\rho(0)= \rho_S(0) \otimes \rho_B(0)$, with $\rho_S(0)$ a generic initial density matrix of the system. This condition can be achieved letting the bath to equilibrate in a thermal state  before preparing the system in the initial condition $\rho_S(0)$ and immediately after switching on the coupling between the two. 
We suppose that at time $t_0=0$ the first projective measurement on the bath degrees of freedom gives the outcome $E_1$ with probability $P(E_1)$.
Then the whole system evolves and after a time interval $t$ the second projective measurement gives  the outcome $E_2$. This second measurement is weighted by the conditional probability $P(E_2;E_1, t)$.
 Denoting with $\varepsilon=E_2-E_1$ the variation of the reservoir energy and with 
 \be
 P(\varepsilon , t)=\sum_{E_1,E_2}\delta(E_2-E_1 -\varepsilon) P(E_1)P(E_2;E_1,t)`~,
 \ee 
 the probability to get $\varepsilon$ from the double measurement in the time interval $t$, the characteristic function can be written as \cite{gasparinetti2014heat, carreganjp}
 \be
 \label{eq:gdef}
 G_\nu (t) = \int_{-\infty}^{+\infty}d \varepsilon e^{i\varepsilon\nu}P(\varepsilon,t)~.
 \ee
 After some simple manipulations (see Ref. \cite{gasparinetti2014heat} for details), the characteristic function can be written as 
 \be
 \label{eq:mgf}
 G_\nu(t) = {\rm Tr}\big[U^\dagger (t) e^{i \nu H_B}U(t)e^{- i \nu H_B}\rho (0)\big]~.
 \ee
 In the above expression $U(t)= \exp{(- i t H)}$ represents the usual time evolution operator from an initial time $t=0$ to time $t$.
 The characteristic function describes all statistical properties of the variation of the reservoir energy and is often called the moment generating function (MGF).
  Indeed the $n$-th derivative of the MGF evaluated at $\nu=0$ gives the $n$th moment of the statistical distribution    
\be
\label{eq:moments}
\langle {\varepsilon^n(t)}\rangle = \left.(-i)^n \frac{d^n }{d \nu^n}G_\nu(  t)\right|_{\nu=0} \, ,
\ee
the symbol $\langle \dots \rangle$ denotes the quantum statistical average.
Hereafter, we focus on  the first moment $\langle \varepsilon(t)\rangle $ corresponding to  the quantum average of energy dissipated by the system into the reservoir  in the time interval $t$. This corresponds to the average energy absorbed (or emitted) by the reservoir $\langle \varepsilon(t)\rangle >0$ ($\langle \varepsilon(t)\rangle <0$).
In particular, in the following we are interested in the temporal dynamics  of the quantum average power $\langle P(t)\rangle $, defined as the time derivative of the quantum average of the energy exchange $\langle \varepsilon(t) \rangle $, i.e.
\be
\langle P(t)\rangle \equiv \frac{ d \langle \varepsilon(t)\rangle }{d t} = -i \frac{d}{d t} \frac{d G_\nu (t)}{d \nu}\big|_{\nu=0}~.
\label{eq:powerdef}
\ee
 This corresponds to the instant power associated to the energy exchange in the TMP.
To proceed in the evaluation of this quantity, several approaches can be used.
In the following we will sketch the derivation using a generalized master equation approach.
We will show results both in the so-called Born-Markov approximation and in the full-secular approximation. Finally we recall the results of Ref.\cite{carreganjp} obtained using a functional integral approach, where non-Markovian contributions at low temperatures are taken into account.
\subsection{Generalized quantum master equation} 
The MGF $G_\nu (t)$ can be rewritten as 
\be
G_\nu(t) = {\rm Tr}\big[\rho^{\nu}(t)\big]={\rm Tr}\big[U_{\nu/2}(t)\rho(0)U_{-\nu/2}^\dagger (t)\big]~,
\ee
where $\rho^\nu (t)$ denotes the generalized total density matrix, the ${\rm Tr}[\dots ]$ is the trace over all degrees of freedom, and $U_\nu (t) = e^{i \nu H_B}U(t)e^{-i \nu H_B}$ is a generalized time evolution operator.

At this point one can write a generalized master equation for the evolution of $\partial \rho^\nu(t) /\partial t$, which reads
\be
\dot{\rho}^{\nu}(t)=-i [H_S,\rho^{\nu}(t)] - i [H_B,\rho^{\nu}(t)] - i [H_I^{\nu},\rho^{\nu}(t)]~,
\ee
where $H_I^{\nu}=e^{i \nu H_B}H_I e^{-i \nu H_B}$.
{ In the interaction picture with respect to the stationary Hamiltonian $H_S + H_B$ any operator $O$ transforms as $O'(t) = e^{i (H_B+ H_S) t}  O e^{- i (H_B+H_S) t}$.} We thus have 
\be
\dot{\rho}^{\nu ' }(t) = -i [H_I ^{\nu '}(t), \rho^{\nu '}(t)]~.
\ee
{ From here on, we follow the standard procedure to obtain the master equation \cite{gasparinetti2014heat, breuer, mandel}.
Under the assumption of weak coupling, first we expand to the second order in the system-bath coupling.
Then, we implement the so-called Born-Markov approximation, which implies that $\rho(t)=\rho_S(t)\otimes \rho_B(0)$, and essentially the Markov approximation which neglects all memory effects of the reservoir. In this scheme we implicitly assume that the time correlation of the reservoir $\tau_B$ are very short with respect to the dissipative dynamics of the system. Consequently the dynamics at short times is valid only when $t\gg \tau_B$ \cite{breuer}. 
Tracing out all bath degrees of freedom we obtain a generalized quantum master equation in terms of the reduced density matrix of the system}
\beq
 \dot{\rho_S}^{\nu '}(t) &=&- \int_0^{+ \infty} d\tau  g(\tau) S'(t) S'(t-\tau) \rho_S^{\nu '}(t) \nonumber \\
&&  + \int_0^{+ \infty}d\tau g(-\nu + \tau)  S'(t-\tau) \rho_S ^{\nu '}(t) S'(t)\nonumber \\
&& +\int_0^{+\infty} d\tau  g(-\nu - \tau) S'(t)  \rho_S^{\nu '}(t) S'(t-\tau)\nonumber \\
&& -\int_0^{+\infty} d\tau g(-\tau) \rho_S ^{\nu '}(t) S'(t-\tau)  S'(t)~.
\label{eq:gqm1}
\eeq
 Note that all integrals in the above equation are taken from $0$ to $\infty$ (because of the Markov approximation) and  $S'(-\tau)$ stands for $\tilde{\sigma}'_z(-\tau)$, which corresponds to the time evolution of the operator coupling the bath to the TSS in the interaction picture. 
 In the above equation we have introduced the correlation function for a bosonic bath
 \be
 g(t)= \int_0^\infty d\omega J(\omega) \left[ \cos (\omega t) \coth(\beta\omega/2) - i \sin (\omega t) \right]~,
 \ee
 
 where $J(\omega)$ is the spectral function of the bath.
 For sake of simplicity, we consider {\it only} the case of Ohmic dissipation, in the scaling limit, where $J(\omega)=K\omega e^{-\omega/\omega_{\rm c}}$, where $\omega_{\rm c}$ is an high energy cut-off and $K$ represents the system-bath coupling strength.
 
 {It is convenient to write equation (\ref{eq:gqm1}) in the basis of the eigenstates of the system Hamiltonian $H_S$.
We thus introduce the transformation $v= e^{- i \sigma_y \theta}$ with $\theta=\arccos(\epsilon/\sqrt{\Delta^2 + \epsilon^2})$ which diagonalizes $H_S$ as $\tilde{H}_S = v H_s v^\dagger$ with eigenvectors $|\psi_1\rangle $, $|\psi_2\rangle$ and eigenvalues $\pm \Omega$ with $\Omega = \sqrt{\Delta^2 + \epsilon^2}$. In the rotated basis one has
}
\be
\tilde{\sigma}_z =v\sigma_z v^\dagger=\cos (\theta)\sigma_z - \sin (\theta )\sigma_x
\label{eq:sigma_z}
\ee
where $\cos(\theta)=\epsilon/\Omega$ and $\sin(\theta)=\Delta /\Omega$. We thus have

\be
S'(t)=\tilde{\sigma}'_z(t) = \cos (\theta) \sigma _z - \sigma _- \sin (\theta)  e^{-i \Omega t } -\sigma _+ \sin( \theta) e^{i \Omega t}~.
\ee

The full master equation, reduced now in terms of the generalized density matrix of the system $\rho_S^{\nu '}(t)$ in the interaction picture can be conveniently cast in a matrix form as
\be
\label{eq:gm2}
\dot{\rho}_S^{\nu ' } = {\bf M}_\nu \rho_S^{\nu '}~,
\label{eq:diffe_eq}
\ee
 where ${\bf M}_\nu$ is a local kernel which describes the evolution of $\rho_S^\nu$.
 The matrix elements of $\rho_S^\nu$ are conveniently expressed in the basis which diagonalize the system Hamiltonian $H_S$
 \be
 \left[\rho_S^{\nu '} \right]_{\alpha , \beta}=\langle \psi_\alpha | \rho_S^{\nu '} |\psi_\beta \rangle~,
 \ee
 with $\alpha, \beta = 1,2$.\\
{ By solving the full generalized dynamics in (\ref{eq:gqm1}), we could, in principle, obtain the whole statistics of the dissipated energy.
Here, we restrict the discussion to the simple problem of determining the behavior of the first moment, i.e., the dissipated energy.\\
The main advantage of considering directly the quantum average power $\langle P(t)\rangle$  is directly related to the structure of the differential equations (\ref{eq:diffe_eq}).
It is sufficient to calculate the system trace 
}
\be
\Sigma_\nu =\sum_{\alpha=1,2} \left[\dot{\rho}_S^\nu \right]_{\alpha, \alpha}= {\rm Tr}_S \big[\dot{\rho}_S^\nu(t) \big]~,
\ee
since 
\be
\langle P(t)\rangle^{(BM)} =-i\frac{\partial \Sigma_\nu}{\partial \nu}\big|_{\nu=0}.
\ee
{ It is worth to note that on the right-hand side of the equation appear only terms which depend on the generalized density matrix of the TSS.
Therefore, we have reduced the initial problem of solving the full dynamics, to the more manageable one of solving a standard master equation for the reduced system dynamics.
For the latter, a large number of analytical and numerical approaches have been developed \cite{schmidt2014, breuer, mandel, orth2013nonperturbative, henriet2014quantum, bulla2003numerical, keil2001real}.
In the specific case of an undriven TSS, after some simple algebra, we arrive at
\be
\label{eq:powermarkov}
\langle P(t)\rangle^{({\rm BM})} = \pi K  \Delta \sin(\theta) \Omega \left(\langle {\sigma_x(t) }\rangle  \coth \left(\frac{\beta  \Omega}{2}\right)-\sin (\theta) \right)~,
\ee
which is the result for the quantum average power in the Born-Markov scheme, as indicated by the index ${\rm BM}$. We stress that in obtaining this result we have kept all the fast oscillating terms in all matrix elements.
In the above expression $\sin(\theta)=\Delta/\Omega$ [see Eq. (\ref{eq:sigma_z})] and $\langle \sigma_x (t) \rangle $ represents the time evolution of the average coherence function of the TSS~\cite{carreganjp, weiss1999quantum}.

\subsection{Full secular approximation}

Another widely used approximation scheme, known as full secular approximation, can be considered. This is a further approximation in the Born-Markov scheme, which greatly simplyfies the calculation. This consists in a full decoupling of the diagonal and the off-diagonal elements of the differential equation in (\ref{eq:gqm1}).
This can be done simply by taking $\exp{(i \Omega t)}\to 0$ in the expressions for the diagonal elements $[\dot{\rho}_S^\nu ]_{1,1}$ and $[\dot{\rho}_S^\nu ]_{2,2}$ of (\ref{eq:gm2}).
This is valid under the assumption that $\Omega$ is faster than any other decoherence rates that appear in the generalized quantum master equation, and any fast oscillating term of the form $e^{i  \Omega t}$ averages to zero.
Under this hypothesis, one can analytically calculate the quantum average power $\langle P(t)\rangle$ following the same lines described in the previous section.
In this regime one obtains
 \beq
 \label{eq:powersecular}
 \langle P(t)\rangle^{({\rm FS})} &=& \pi K  \Omega^2 \sin ^2(\theta ) \nonumber \\ 
 && \times \left[\coth \left(\frac{\beta 
   \Omega}{2}\right) \left(\langle {\sigma_x(t) }\rangle \sin (\theta) +\langle{\sigma_z(t)}\rangle \cos ( \theta)\right)-1\right]~,
   \eeq
   where $\langle \sigma_z (t) \rangle $ describes the time evolution of the average population function~\cite{carreganjp, weiss1999quantum, grifoni1996exact, grifoni1997coherences}. It is worth to recall that both approaches, the Born-Markov (${\rm BM}$) and the full secular (${\rm FS}$), rely on the assumption of weak coupling between the bath and the system, with the coupling strength $K\ll 1$.
Note that both approaches based on the generalized master equation does not reproduce the correct dynamics for very short time, while considering $t \ll \tau_B$, with $\tau_B$ the characteristic correlation time of the reservoir.

 \subsection{Functional integral approach}
A different route towards the evaluation of the energy exchange between the thermal bath and the TSS is represented by the functional integral approach for quantum dissipative systems \cite{weiss1999quantum, feynman1963theory, caldeira1983path, leggett1987dynamics, ingold2002path, chen1989}.
 In Ref \cite{carreganjp} it was developed a formalism to deal with a generic quantum system coupled with a reservoir, by means of well-known path-integral dissipative  techniques.
 In essence, this is a generalization of the Feynman-Vernon functional for a quantum dissipative system for the inspection of energy exchange processes. In particular, an influence functional was constructed to evaluate all statistical properties of the variation of energy of a reservoir coupled to a generic quantum system.\\
The starting point is the expression of the characteristic function $G_\nu(t)$ in (\ref{eq:gdef}).
By taking advantage of the gaussian properties of the set of the harmonic oscillators, one can trace out all the degrees of freedom of the reservoir and consider the reduced dynamics of the quantum system, ending up with a functional integral which embodies all the effect of quantum dissipation and energy exchange between the reservoir and the system.
 The details of this procedure are presented in Ref.~\cite{carreganjp}.
 In the case of a two state system the characteristic function takes the form
 \be
G_\nu(t) = \int d\eta_i \,\bra{  \eta_i} \rho_{S}(0) \ket{\eta_i} \int d\eta_f   \!\!\!
\int \! \! \D \eta \int \! \! \!\D \xi\;
{e}^{i\,S_{ S}[\eta,\xi]}\;  {\cal F}_{{\rm FV}}[\eta,\xi]\cdot {e^{i \Delta \Phi^{(\nu)}[\eta,\xi]}}~, 
\ee
where integrations run over all possible initial and final states of the TSS.
In the above expression ${ S}_S$ represents the action of the system, while ${\cal F}_{{\rm FV}}$ is the usual Feynman-Vernon functional of quantum dissipative TSS, see for example \cite{carreganjp, weiss1999quantum} (see Ref.\cite{carreganjp} for details). The new functional $\Delta \Phi^{(\nu)}[\eta,\xi]$ carries formally all statistical information of energy exchange.\\
Using this formalism one in principle can access to all moments of the statistical distribution and is valid for general linear dissipation and any type of memory-friction, and for different strength of system-bath interaction. Moreover, it does not depend on the specific nature of the quantum system and it can be used as a starting point for different powerful numerical and analytical  methods \cite{weiss1999quantum, grifoni1998driven, orth2013nonperturbative, henriet2014quantum, bulla2003numerical, keil2001real, sassetti1990universality}.\\
Considering the paradigmatic case of linear spectral function $J(\omega ) = K \omega e^{-\omega/\omega_{\rm c}}$ (the Ohmic case of interest in this work) and weak coupling $K\ll 1$, it is possible to get a compact analytical expression, exact at the lowest order of $K$, for the time-evolution of the quantum average power $\langle P(t)\rangle$.
Here, we report only the final result,as obtained in Ref.~\cite{carreganjp}
\beq
\label{eq:powerpath}
\langle P(t)\rangle^{({\rm PI})} &=& - \frac{\Delta}{2 \pi} \int_0^t d\tau \Re e [g(\tau)] \Big( \,\langle\sigma_x(t-\tau)\rangle_{\rm s}  \,\langle\sigma_z(\tau)\rangle_{\rm s} -
\langle\sigma_z(t-\tau)\rangle_{\rm a}\, \langle\sigma_x(\tau)\rangle_{\rm a} \Big) \nonumber \\
&& +\frac{\Delta }{2 \pi}\int_0^t d\tau \Re e [g(\tau)]
\Big(\langle\sigma_x(t-\tau)\rangle_{\rm a}  \,\langle\sigma_z(\tau)\rangle_{\rm s} -
\langle\sigma_z(t-\tau)\rangle_{\rm s}\, \langle\sigma_x(\tau)\rangle_{\rm a} \Big) \nonumber \\
&& + \frac{\pi K}{2}\Delta^2.
\eeq
The above quantity is expressed in terms of a time convolution  which involves in the  time segment the population correlation  function $\langle {\sigma_z(t)}\rangle $~\cite{weiss1999quantum, leggett1987dynamics, grifoni1996exact, grifoni1995cooperative}
and the coherence correlation  function $\langle {\sigma_x(t)}\rangle $ of the TSS~\cite{weiss1999quantum, grifoni1997coherences, grifoni1996exact, grifoni1995cooperative}. The average power $\langle P (t)\rangle^{({\rm PI})}$ depends on the polarization correlation function of the reservoir which takes into account possible memory effects. These contributions become relevant and strongly affect the dynamics at low temperatures and short times as we will show below.
The indices $s$ and $a$ which appear in (\ref{eq:powerpath}) indicate that such quantities are symmetric and antisymmetric terms under bias inversion $\epsilon \to -\epsilon$.

\section{Results and discussion}
\label{sec:result} 
In this section we discuss the behaviour of the quantum average power $\langle P(t)\rangle$, comparing the results obtained with the different schemes exposed above.
We underline the main differences between the non-perturbative approach based on the functional integral  and the ones based on the quantum master equation, showing in which regimes they reproduce comparable results or not.
We consider only the case of {\it weak} coupling, with $K \ll1$, where all approaches are potentially valid and can be compared.
 For sake of definiteness, hereafter we fix $K=0.02$.
We also assume an initial condition with the TSS in a diagonal state with probability $p_l - p_r=1$.
Generalization to other cases is straightforward~\cite{carreganjp, weiss1999quantum}.\\
In general, all approaches reproduce the same behaviour for the quantum average power at high temperatures $T\gg \Omega = \sqrt{\Delta^2+\epsilon^2}$, while marked differences can be found in the opposite regime where the temperature is of the same order or lower than $\Omega$.\\
Since the  quantum average power $\langle P(t)\rangle$, in all approaches, is expressed in terms of the population and the coherence correlation function $\langle \sigma_z (t)\rangle $ and $\langle \sigma_x (t)\rangle $ respectively [see (\ref{eq:powermarkov}), (\ref{eq:powersecular}) and (\ref{eq:powerpath})]. We write their explicit for \cite{carreganjp, weiss1999quantum, grifoni1998driven, gorlich1989low}. One has
\be
\langle \sigma_i (t)\rangle = \langle \sigma_i (t)\rangle_s + \langle \sigma_i (t)\rangle_a \qquad i=x,z~
\ee
    where the symmetric and antisymmetric (with respect to bias inversion) contributions are given by \cite{weiss1999quantum}
    \be
    \label{eq:szs}
    \langle \sigma_ z (t) \rangle _s = \frac{\epsilon^2}{\Omega^2}e^{-\gamma_r t} + \frac{\Delta^2}{\Omega^2}\cos(\Omega t)e^{-\gamma t}~,
    \ee
    \be
    \langle \sigma_z (t)\rangle_a= \frac{\epsilon \tanh (\beta \Omega/2)}{\Omega}\left(1-e^{-\gamma_r t}\right)~,
    \ee
    \be
    \langle \sigma_x(t)\rangle_s =\frac{\Delta \tanh(\beta \Omega/2)}{\Omega}\left(1-e^{-\gamma_r t}\right)~,
    \ee
    and
    \be
    \langle \sigma_x(t)\rangle_a= \frac{\epsilon \Delta}{\Omega^2}\left[e^{-\gamma_r t} - \cos(\Omega t) e^{-\gamma t}\right]~.
    \ee
     In the above equations we have introduced the relaxation $\gamma_r$ and damping  $\gamma$ rates. These are connected by the so-called  Vieta relations \cite{weiss1999quantum}
     \beq
    \label{eq:vieta}
    \gamma + \gamma_r &=& \frac{4\pi K}{\beta}\nonumber \\
    \gamma_r (\gamma^2 + \Omega^2) &=&\frac{2\pi K\Delta^2}{\beta} \nonumber \\ 
\gamma^2 + 2 \gamma \gamma_r + \Omega^2 &=&\left(\frac{2\pi K}{\beta}\right)^2 + \epsilon^2 + \Delta^2~.
\eeq
In the {\it weak} coupling limit $K\ll 1$ the relaxation rate is given by \cite{weiss1999quantum}
     \be
     \gamma_r = \pi K \frac{\Delta}{\Omega}\coth \left(\frac{\beta \Omega}{2}\right)~.
     \ee
  With these ingredients one can calculate the time evolution of the quantum average power given in equations (\ref{eq:powermarkov}), (\ref{eq:powersecular}), and (\ref{eq:powerpath}).\\   
     In Figure \ref{fig:1} we show the quantum average power as a function of time, for a fixed external constant bias $\epsilon=\Delta$. Different curves are obtained considering different approximation schemes.
The choosen initial conditions and the finite external bias are responsible of  a finite value of $\langle P(t)\rangle$.
For long times, the quantum average power vanishes, in all approaches, when the whole system reaches equilibrium, as one would expect for the case of constant bias.
This feature is clearly visible in Figure \ref{fig:1} in all curves for sufficient long time.
On the contrary, the transient dynamic evolution towards equilibrium shows different features for the three approaches considered here.
In particular, the master equation approach in the full secular approximation ${\rm FS}$ (dotted line) is able to capture {\it only} the exponential decay towards equilibrium.
The quantum average power $\langle P(t)\rangle$ shows an oscillating behaviour if one consider the path integral approach ${\rm PI}$ (solid line) or the Born-Markov scheme ${\rm BM}$ (dashed line).
In passing we note that these different behaviours (relaxation decay or oscillations) are analogous to the ones obtained in similar approaches for the population dynamics of the dissipative TSS \cite{weiss1999quantum, grifoni1996exact, grifoni1995cooperative}
\begin{figure}[ht]
\centering
\includegraphics[scale=0.60]{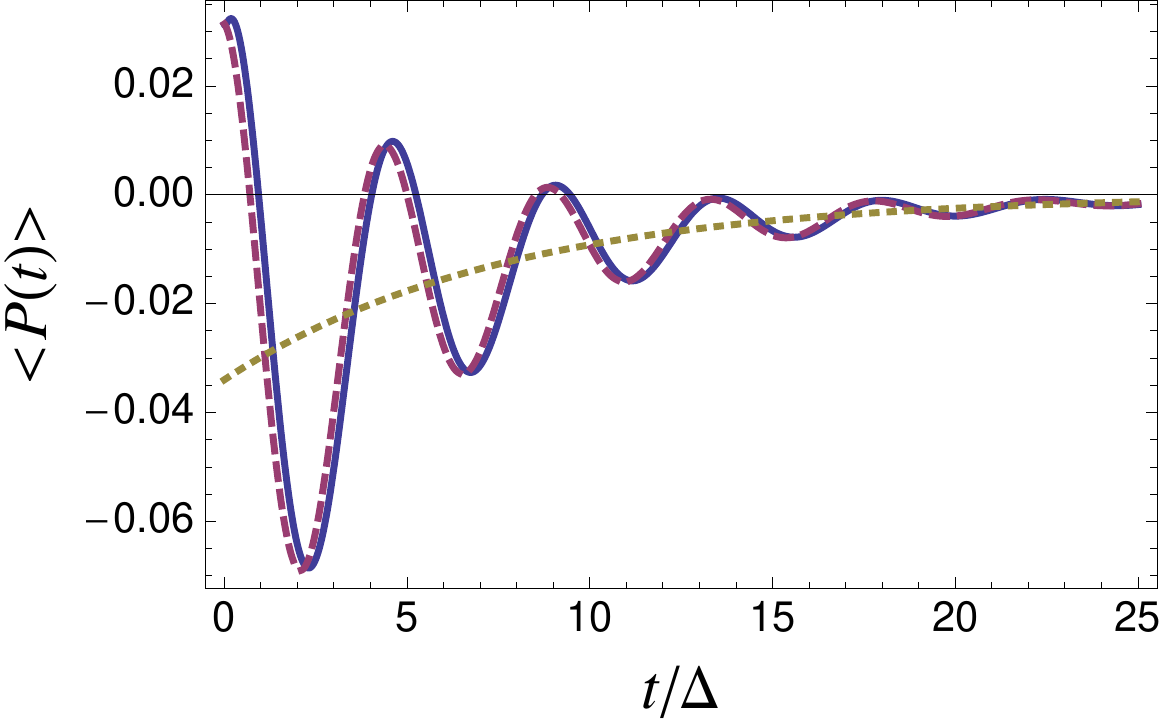}
\includegraphics[scale=0.60]{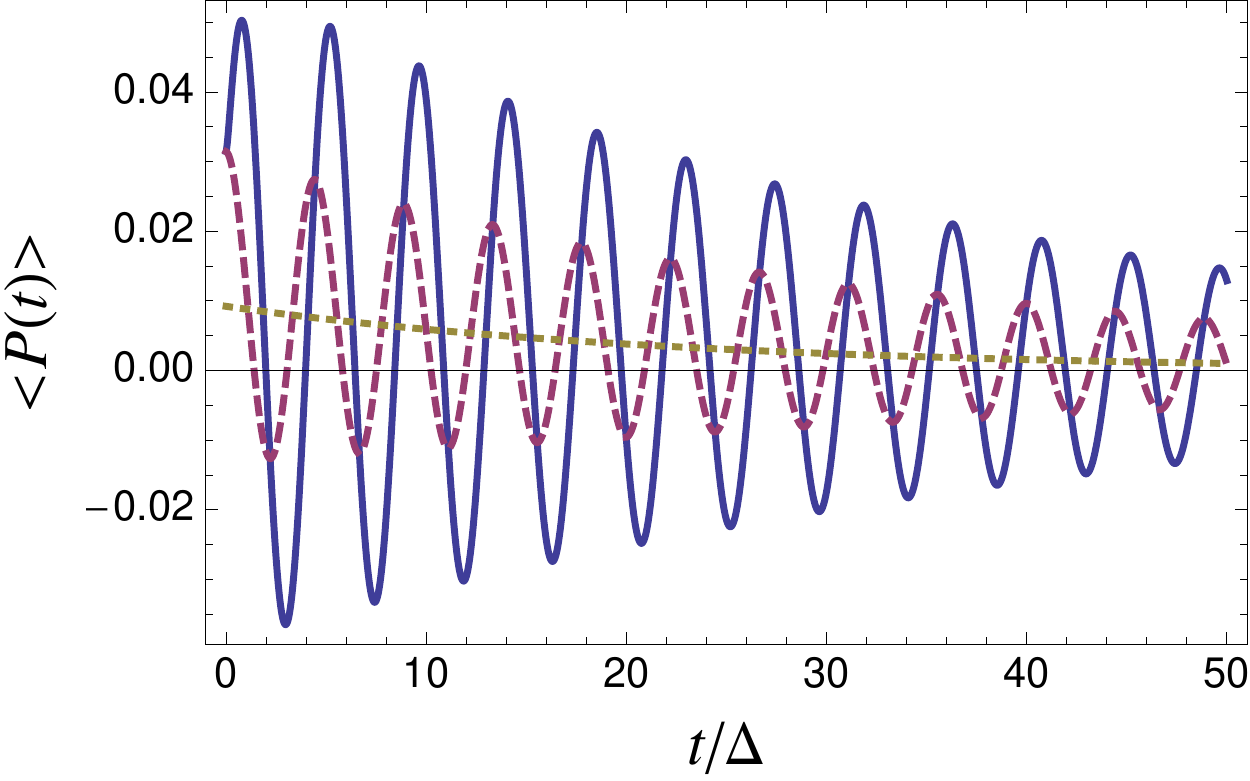}
\caption{(Colour online). Quantum average power $\langle P(t)\rangle $  in units of $\Delta^2$ as a function of time. 
Different curves refer to different approximation schemes: Born-Markov ${\rm BM}$ (dashed line), full secular $FS$ (dotted line), and  path-integral $PI$ (solid line) approach.
The parameters are $K=0.02$ and $\epsilon =1$.
The bath temperature is $T=2$ (left panel) and $T=0.05$ (right panel).
In both cases, times, frequencies and temperature are scaled with $\Delta$.
 }
\label{fig:1}
\end{figure}
From the left panel of Figure \ref{fig:1}, where the temperature is set to $T=2\Delta$, it is evident that the master equation approach without secular approximation give quite accurate results, with a very similar oscillating behaviour of the PI curve. Note that the ${\rm FS}$ curve reproduces {\it only} the relaxation dynamics, with the same decaying behaviour but with no oscillations.
Differences between the ${\rm BM}$ and ${\rm PI}$ approaches become evident while lowering the temperature $T\leq \Omega$, where clearly non-Markovian effects are relevant. The latter play an important role at low temperature and are captured {\it only} by the path integral approach \cite{carreganjp}.
These deviations are shown in the right panel of Figure \ref{fig:1}, where the temperature is set to $T=0.05 \Delta$.

The left panel of Figure \ref{fig:2} shows the behavior of the average dissipated energy $\langle \varepsilon(t)\rangle $ as a function of the bath temperature $T$. This quantity is obtained by direct integration of $\langle P(t)\rangle $ and the observation time $t$ is now fixed to $10/\Delta$.
Here, the overall qualitative behaviour of $\langle \varepsilon(t)\rangle$ is well described within all the three approaches, (see the different curves in the Figure). 

It is possible to analytically demonstrate that in the high temperature regime all curves collapse, giving the same result for $\langle \varepsilon(t)\rangle$ (although the asymptotic behaviour is not yet reached in the left panel of Figure \ref{fig:2}).  At low temperature instead, the ${\rm BM}$ and ${\rm FS}$ give very similar results, with visible differences with the ${\rm PI}$ one.
Notice that differences between the three approaches vanish for long integration time, since the oscillations of the dissipated quantum average power (see Figure \ref{fig:1}) average out.
Therefore marked and strong differences can be found in the low temperature regime and for sufficient short times.
This analysis confirm that the more rigorous ${\rm PI}$ approach could give very different results with respect to simpler approaches such as the one based on different approximations with generalized master equation.
In essence this is due to the fact that non-Markovian and coherent contributions are properly taken into account with the functional integral approach.
However in specific regimes the ${\rm BM}$, and also the ${\rm FS}$, schemes can reproduce very similar results and are thus well suited for the evaluation of such quantities.

{ The right panel of Figure \ref{fig:2} gives a quantitative information about deviation between the different approaches.
We have chosen to focus on the solutions that preserve the non-Markovian dynamics, i.e. the path integral, and on the Born-Markov approaches, and studied them as a function of the temperature $T/\Delta$ and the external bias $\epsilon_0/\Delta $.
As distance estimator we use the difference between $\langle \varepsilon (t) \rangle^{({\rm PI})}$ and $\langle \varepsilon (t)\rangle^{({\rm BM})}$ normalized to $\Delta$.
In the density plot the observation time is fixed to $t=10/\Delta$ as in the other panel.
Light and dark colours represent large and small deviations, respectively.
As we can see, the differences are maximal for low temperature when they reach $0.02 \Delta$ and show an oscillating behavior as a function of  $\epsilon_0/\Delta $.
}


\begin{figure}[ht]
\centering
\includegraphics[scale=0.60]{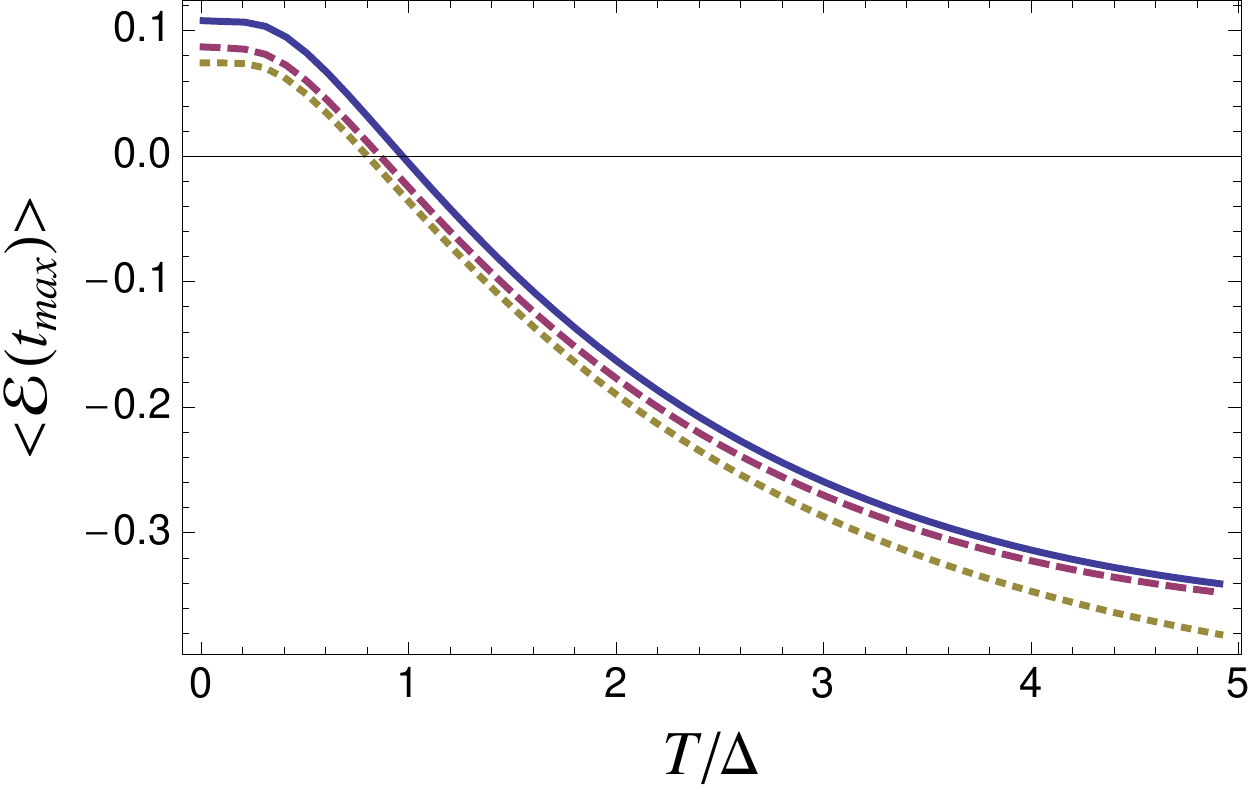}
\includegraphics[scale=0.55]{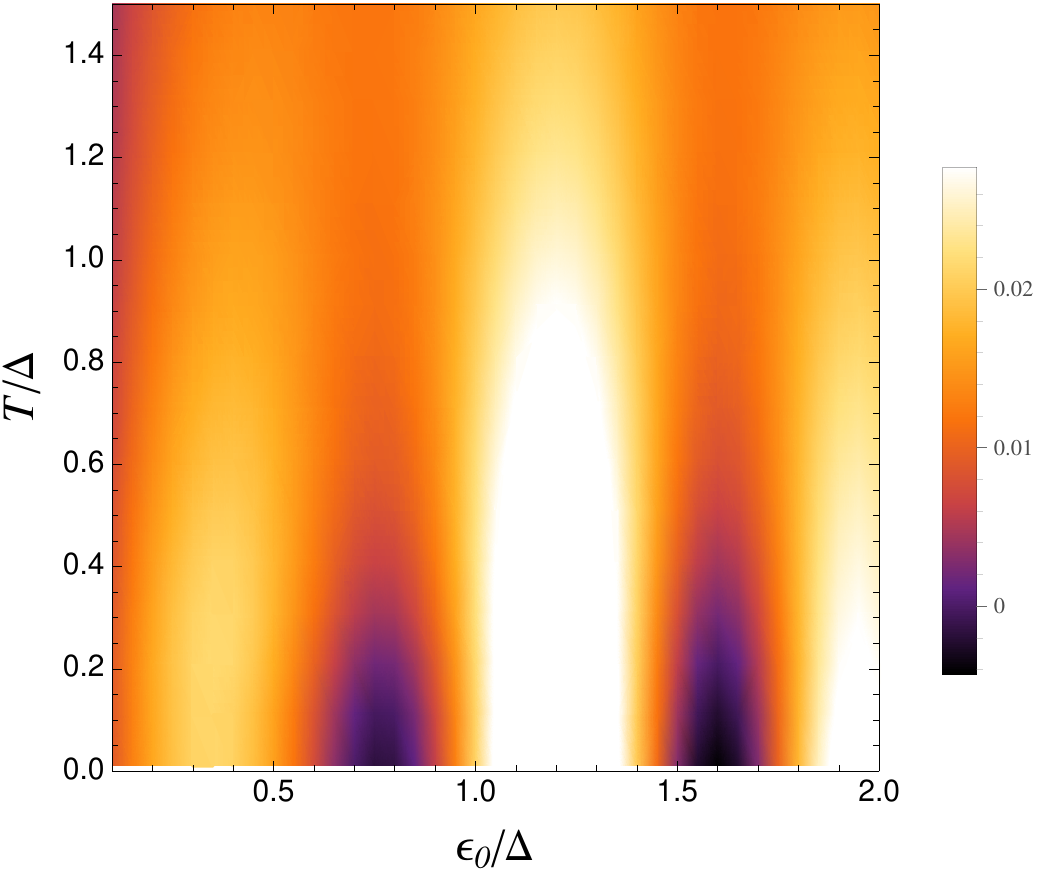}
\caption{(Colours online) The left panel shows the integrated quantum average power $\langle \varepsilon(t)\rangle $, or dissipated energy, (in units of $\Delta$) as a function of temperature $T$ . Different curves refer to different approximation schemes: Born-Markov $BM$ (dashed line), full secular $FS$ (dotted line), and Path Integral $PI$ (solid line) approach.
The other parameters are $K=0.02$ and $\epsilon=1$. Times, frequencies and temperature are scaled with $\Delta$. 
The density plot in the right panel shows the difference $\langle \varepsilon(t)\rangle^{({\rm PI})} - \langle \varepsilon(t)\rangle^{({\rm BM})}$ of the average dissipated energy (in units of $\Delta$) between the PI and the BM approaches as a function of $\epsilon_0/\Delta$ and $T/\Delta$. Large deviations are marked with lighter colour, while black regions represent the case of vanishing deviations with $\langle \varepsilon (t)\rangle^{({\rm PI})} \approx\langle \varepsilon(t)\rangle^{({\rm BM})}$.
In both panels the measurement time is fixed to $t_{max}=10/\Delta$ in order to enhance the visibility of the nonn-Markovian contributions.
}
\label{fig:2}
\end{figure}

\section{Conclusions}
We have investigated the energy exchange between a two-state system, subjected to a constant external bias, and a thermal reservoir. In particular we have focussed on the variation of energy $\langle \varepsilon (t)\rangle$ of the reservoir in a given time interval in the case of weak system-bath coupling.
This quantity can be measured, for instance, by means of a protocol, based on two subsequent projective measures of the bath energy.
The temporal dependence of the time derivative of this quantity $\langle P(t)\rangle$ shows different transient behavior depending on the approximation and the approaches used to solve
the dynamics.
In particular, we have compared the results obtained by solving a master equation with and without the secular approximation and the functional integral approach.
The transient behavior of $\langle P(t)\rangle$ is correctly described by the  functional integral formulation and the Born-Markov master equation without
the secular approximation.
We have shown that the solution with secular approximation is able to catch {\it only} the coarse-grained dynamics, reproducing only an exponential decay for the quantum average power.
At low temperature, the difference between the path integral and master equation approach increases as we have discussed, as a consequence of the increasing importance of non-Markovian effects in that regimes.

\section*{Acknowledgments}
We thank U. Weiss for useful discussions. We acknowledge the support of the MIUR-FIRB2012 - Project HybridNanoDev (Grant  No.RBFR1236VV), EU FP7/2007-2013 under REA grant agreement no 630925 -- COHEAT, MIUR-FIRB2013 -- Project Coca (Grant
No.~RBFR1379UX), and the COST Action MP1209.
A.B. also acknowledges support from STM 2015 from CNR and Victoria University of Wellington where this work was partially done.

\end{document}